\documentclass[iop]{emulateapj}

\usepackage{graphicx,amsmath,amsfonts,amssymb,subfigure,hyperref}

\newcommand{\kms}{\,~km~s$^{-1}$}

\def\rsun{\ifmmode {\rm R}_{\mathord\odot}\else $R_{\mathord\odot}$\fi}
\def\msun{\ifmmode {\rm M}_{\mathord\odot}\else $M_{\mathord\odot}$\fi}
\def\lsun{\ifmmode {\rm L}_{\mathord\odot}\else $L_{\mathord\odot}$\fi}

\newcommand{\nhtd}{\mbox{\rm NH$_2$D(1$_{1,1}$-1$_{0,1}$)}}
\newcommand{\nthp}{\mbox{\rm N$_2$H$^+$(1-0)}}
\newcommand{\hcn}{\mbox{\rm HCN(1-0)}}
\newcommand{\hcop}{\mbox{\rm  HCO$^+$(1-0)}}
\newcommand{\cs}{\mbox{\rm CS(2-1)}}

\newcommand{\ceo}{\mbox{\rm C$^{18}$O(1-0)}}

\slugcomment{\it Accepted for publication in ApJ Oct 4, 2017}

\begin{document}

\title{A turbulent origin for the complex envelope kinematics in the young low-mass core Per-Bolo 58 }
\author{Mar\'ia Jos\'e Maureira, H\'ector G. \ Arce}
\affil{Astronomy Department, Yale University, New Haven, CT~06511, USA}
\email{mariajose.maureira@yale.edu, hector.arce@yale.edu}

\author{Stella S. R. Offner}
\affil{Department of Astronomy, The University of Texas at Austin, Austin, TX, 78712}

\author{Michael M. Dunham}
\affil{Department of Physics, State University of New York at Fredonia, Fredonia, NY 14063, USA}

\author{Jaime E. Pineda}
\affil{Max-Planck Institute for Extraterrestrial Physics, Giessenbachstrasse 1, 85748 Garching, Germany}

\author{Manuel Fern\'andez-L\'opez}
\affil{Instituto Argentino de Radioastronom´ıa (CCT-La Plata, CONICET; CICPBA), C.C. No. 5, 1894,Villa Elisa, Argentina}

\author{Xuepeng Chen}
\affil{Purple Mountain Observatory, Chinese Academy of Sciences, 2 West Beijing Road, Nanjing 210008, China}

\and

\author{Diego Mardones}
\affil{Departamento de Astronom´ıa, Universidad de Chile, Casilla 36-D, Santiago, Chile}

\begin{abstract}

We use CARMA 3mm continuum and molecular lines (NH$_2$D, N$_2$H$^+$, HCO$^+$, HCN and CS) at $\sim$1000 au resolution to characterize the structure and kinematics of the envelope surrounding the deeply embedded first core candidate Per-Bolo 58. The line profile of the observed species shows two distinct peaks separated by 0.4-0.6 \kms, most likely arising from two different optically thin velocity components rather than the product of self-absorption in an optically thick line. The two velocity components, each with a mass of $\sim$0.5-0.6 M$_{\odot}$, overlap spatially at the position of the continuum emission, and produce a general gradient along the outflow direction. We investigate whether these observations are consistent with infall in a turbulent and magnetized envelope. We compare the morphology and spectra of the \nthp\ with synthetic observations of an MHD simulation that considers the collapse of an isolated core that is initially perturbed with a turbulent field. The proposed model matches the data in the production of two velocity components, traced by the isolated hyperfine line of the \nthp\ spectra and shows a general agreement in morphology and velocity field. We also use large maps of the region to compare the kinematics of the core with that of the surrounding large-scale filamentary structure and find that accretion from the large-scale filament could also explain the complex kinematics exhibited by this young dense core.

\end{abstract}


\section{Introduction}
\label{sec_intro}

Low mass protostars form from the gravitational collapse of dense cores inside molecular clouds (\citealt{1989BensonSurvey,2002CaselliDense,1969LarsonNumerical,1987ShuStar}). According to theory, the details of this process vary with the initial conditions in the core such as the presence of magnetic fields, rotation, and turbulence. For instance, magnetic fields affect the infall of material towards the center of the core; the accretion of material occurs preferentially along the direction of the magnetic field, which produces a more flattened shape of the dense material at the center of the core (\citealt{1976MouschoviasNon,2014BateCollapse,2014LiLink}). Also, magnetic fields play part in the production of outflows before and after protostar formation (\citealt{2011MatsumotoProtostellar, 2013TomidaRadiation, 2014MachidaProtostellar, 2014BateCollapse,2015TomidaRadiation}).
Rotation also contributes to a more flattened geometry of the dense material \citep{2010BateCollapse}, and later on to the formation of the circumstellar disk \citep{2011MachidaOrigin}. Turbulence has been shown to break the symmetry of the collapse. In simulations of turbulent cores, the gas quickly becomes clumpy and the infalling gas may flow through dense streams or narrow channels which can be initially aligned with the magnetic field (\citealt{2011MatsumotoProtostellar,2011SmithQuantification,2012SmithLine,2015SeifriedAccretion}). The overall morphology of turbulent cores in simulations is more consistent with prolate, triaxial or filamentary shapes \citep{2008OffnerDriven,2011SmithQuantification}. If the magnetic field is weak with respect to the turbulence, the direction of field lines at 1000 au scales can be different from the initial field direction at larger scales (\citealt{2011MatsumotoProtostellar}). Turbulence can also imprint bulk rotation in the gas, in which case the rotation axis can be misaligned with the magnetic field direction. In this last scenario, more massive disks are allowed to form, as well as outflows that are not aligned with the surrounding magnetic field (\citealt{2000BurkertTurbulent,2010CiardiOutflows,2011MatsumotoProtostellar,2013JoosInfluence,2017LeeSynthetic}).

Molecular lines tracing the envelope of young dense cores provide information on the gas kinematics and thus, are useful for testing the effects of turbulence, magnetic field, and rotation in collapsing cores. For instance, a transition from more turbulent edges to more quiescent dense regions is expected if cores are formed from turbulent gas (e.g. \citealt{2008OffnerKinematics}). Consistent with this, single-dish observations show that the non-thermal linewidths of optically thin tracers like NH$_3$ and N$_2$H$^{+}$ become coherent and less turbulent towards the denser regions where one or more cores are located (\citealt{1998GoodmanCoherence,2010PinedaDirect,2011HacarDense,2015SeoAmmonia,2017FriesenGreen}).

The kinematics of dense cores, as traced by thin molecular tracers, also show velocity gradients from few 0.1 pc to 1000 au scales (\citealt{1993GoodmanDense,2002CaselliDense,2007ChenOvro,2011TobinComplex,2011PinedaEnigmatic}). Depending on the geometry of the envelope and the direction of the gradient they are typically interpreted as produced by rotation and/or infall. Rotation can produce velocity gradients along the major-axis of a flattened envelope, perpendicular to the outflow axis. Some examples of young sources that show this type of velocity distribution can be found in \cite{2002BellochMolecular,2011TobinComplex,2011TannerDynamics,2013YenUnveiling,2016OyaInfalling,2017MaureiraKinematics}. On the other hand, velocity gradients along the minor-axis of flattened envelopes, viewed close to edge-on, can arise from infall motions (\citealt{1995TorrellesSignatures,1999OhashiCCS,2002CaselliMolecular,2010YenHigh}). In addition, if the envelope is filamentary, as those produced by simulation that include turbulence,  a velocity gradient along the major-axis can be also produced by infall rather than rotational motions (\citealt{2012TobinComplexIII}).\\

Infall and rotation can also be detected using optically thick molecular lines such as \hcn, \hcop\ and \cs. In this case, the shape of the emission line will depend on changes in the kinetic temperature and density (excitation temperature) as well as velocity along the line of sight. A two-peak profile with a more intense blue-shifted peak (also known as blue asymmetric profile) is expected to arise from infall motions in isolated spherically symmetric envelopes in which the density increases toward the center (\citealt{1999EvansPhysical,2011TomisakaObservational}). On the other hand, simulations of cores formed from in a turbulent environment show a more complex picture due to the asymmetries in density and the disordered velocity field, leading to highly variable profiles for the same core as seen along different lines of sight. \cite{2012SmithLine} simulated optically thick and thin tracers toward the center of dense cores embedded in filaments which were in turn formed out of a turbulent molecular cloud. They showed that even though the gas motions were dominated by infall, optically thick tracers can show both blue and red asymmetries in their double-peak profile, depending on the inclination and azimuthal angle used for observing the core. Red-asymmetric profiles appeared in these turbulent collapsing cores due to either the contribution of a filamentary envelope or accretion occurring from only one side.\\

\begin{figure}
\center
\includegraphics[width=0.5\textwidth]{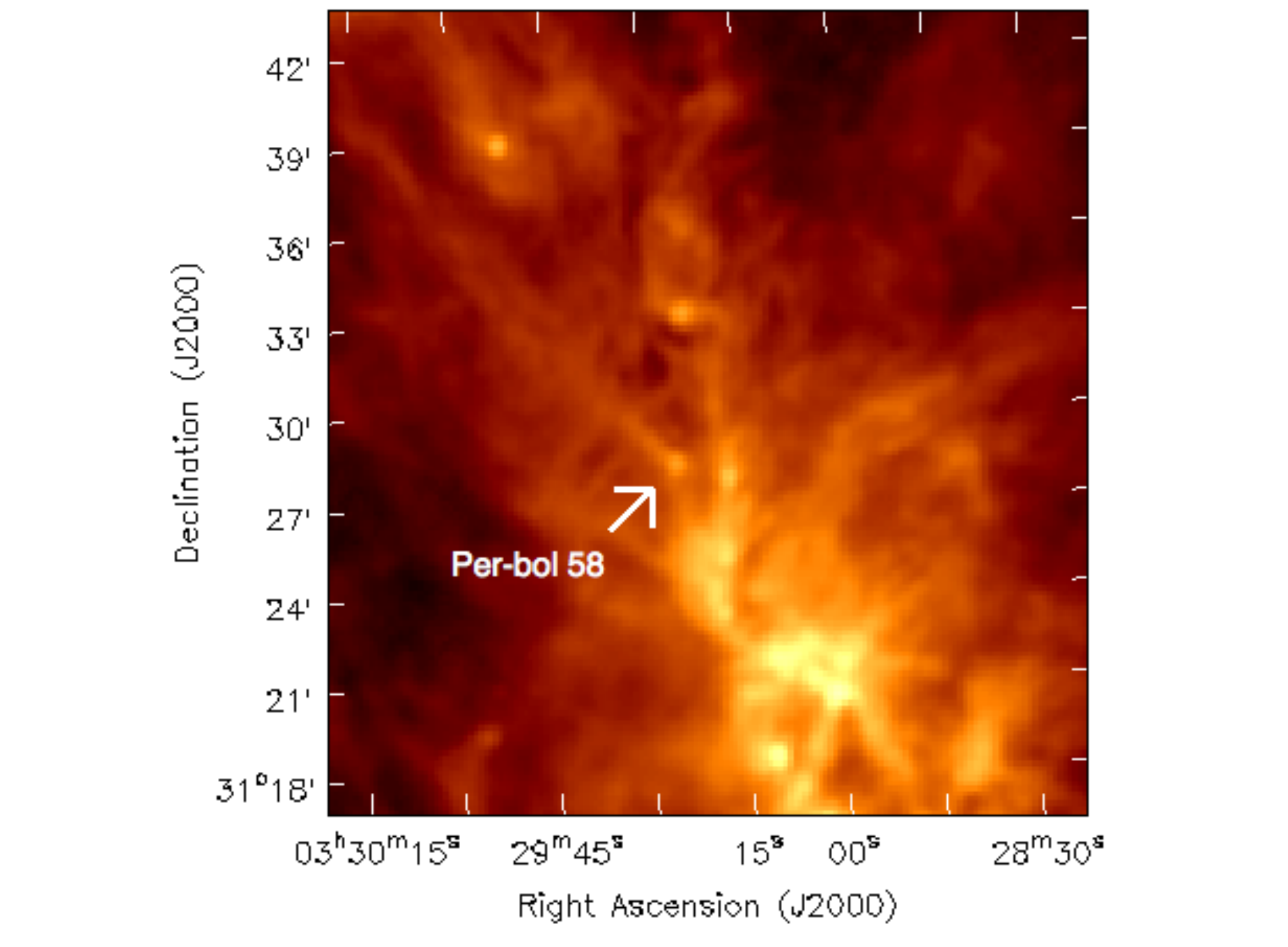}
\caption{Herschel 250 $\mu$m map of the NGC1333 region in Perseus. The white arrow indicates the position of the dense core Per-Bolo 58.\label{fig:herschel}}
\end{figure}

In this work we present CARMA (Combined Array for Research in Millimeter-wave Astronomy) molecular line observations at 1000 au scales of the young dense core Per-Bolo 58, located North of the NGC1333 cluster region in Perseus, at a distance of 230 pc \citep{2008HirotaAstrometry}. Figure~\ref{fig:herschel} shows the location of Per-Bolo 58 in a 250 $\mu$m Herschel map which shows that the dense core is embedded in a large scale filamentary structure. We study Per-Bolo 58 kinematic using optically thin and thick tracers (\nhtd, \nthp, \hcn, \hcop\ and \cs). \\

Per-Bolo 58 is an ideal source for studying the conditions in which the gas is transported towards the core center in young sources; the spectral energy distribution (SED) of this source suggests that the compact source at the center is not a protostar yet, and corresponds instead to a theoretical object called a first hydrostatic core (see below). A first hydrostatic core (a.k.a. FHSC or simply first core) is a transient object ($10^{3}-10^{4}$ years) that reaches quasi-hydrostatic equilibrium at the center of a dense core before the formation of a protostar (\citealt{1969LarsonNumerical}). Unlike protostars, first cores are objects made of molecular hydrogen, with a size of  1-20 au and a central temperature of a few 100 K up to 2000 K. At 2000 K the molecular hydrogen dissociates and a collapse at the inner 0.1 au leads to the formation of a protostar (\citealt{1969LarsonNumerical,1998Masunaga,2006SaigoEvolution,2011MatsumotoProtostellar,2012JoosProtostellar,2013TomidaRadiation,2014BateCollapse,2015TomidaRadiation}). \\

Per-Bolo 58 was identified as a first core candidate by \cite{2010EnochCandidate} based on its SED which shows that the central source has a very low luminosity (L$_{int}\sim0.012$ L$_{\odot}$), lower than the typical Class 0 object and even Very Low Luminosity Objects (VeLLOs). Using this same criteria past studies have reported about nine other first core candidates (e.g. \citealt{2006BellocheEvolutionary,2010ChenL1448,2010EnochCandidate,2011DunhamDetection,2011PinedaEnigmatic,2012ChenSubmillimeter,2012SchneeHow,2012PezzutoHerschel, 2013HuangProbing, 2013MurilloDisentangling}).

The central source in Per-Bolo 58 is driving a bipolar outflow, which was detected using SMA CO(2-1) observations with a resolution of 2.7" \citep{2011DunhamDetection}. The outflow is slow with a characteristic velocity of 2.9 \kms\ and shows a compact morphology with a an opening semi-angle of about 8 degrees for both lobes. This is only partially consistent with several simulations of outflows launched by first cores, which are typically low in velocity ($\lesssim10$ \kms) but also poorly collimated \citep{2011MatsumotoProtostellar, 2013TomidaRadiation, 2014MachidaProtostellar, 2014BateCollapse,2015TomidaRadiation}. In addition, the dynamical time inferred from these outflow observations is $10^4$ years, which is close to the upper limit of the lifetime of first cores, derived in simulations \citep{2010TomidaExposed}. Although the outflow morphology and its estimated lifetime could lead to the straightforward conclusion that Per-bolo 58 is a protostar (and not a first core), studies of the outflow launching mechanism in first cores and very young protostars have shown that outflow collimation depends on the interplay between rotation and the strength of the magnetic field (assuming is initially parallel to the rotation axis), which in some cases could lead to well-collimated jet-like outflows, even in first cores (\citealt{2002TomisakaCollapse,2012PriceCollimated,2012SeifriedMagnetic}). Also, as pointed out by \cite{2011DunhamDetection}, observations providing shorter baselines and/or deeper single-dish data could reveal a wider component in Per-bolo 58's outflow. Observations of this source have thus far shown it is very young, but have not been able to confirm whether it is in the first core stage or if it is a very young (Class 0) protostar.\\

The paper is organized as follows: in Section 2, we describe the observations; in Section 3, we present our general results; in Section 4, we analyze the kinematics; and in Section 5, we discuss the origin of  Per-bolo 58's kinematics. We investigate an envelope scale and a filament scale turbulent scenario. For the former, we compare our observations with MHD simulations of the collapse of an isolated turbulent and magnetized core. In Section 6 we discuss the evolutionary state of this first core candidate. Section 7 corresponds to the summary and conclusions.

\section{Observations}

Per-Bolo 58 was observed with the Combined Array for Research in Millimeter-wave Astronomy (CARMA) between April and August 2012, using the two most compact configurations (D and E) which provided baselines from 8 to 150 m, and thus sensitivity to structures up to 50" (12000 au). The correlator configuration consisted of seven 8 MHz windows with 319 channels ( $\Delta v \sim 0.085$ km/s) for spectral lines and two 500 MHz windows for continuum. The seven narrow windows were centered in the following molecular transitions: \nthp, \cs, \hcop, \hcn, \nhtd, SiO(2-1), and C$^{34}$S(2-1). The first core candidate L1451-mm (\citealt{2017MaureiraKinematics}) was observed together in the same track with Per-Bolo 58. Identical calibrations and imaging procedures were applied to both. See (\citealt{2017MaureiraKinematics}) for more details regarding observations and data reduction.

The synthesized beam size of the maps is $\sim$ 6", which at a distance of the Perseus molecular cloud (230 pc, \citealt{2008HirotaAstrometry}) corresponds to 1380 au. Channels were re-sampled to a resolution of 0.1 \kms during imaging.  Table~\ref{tb:perbol58_maps} lists the array configurations, synthesized beam and RMS of the final Per-Bolo 58 maps. For reference, Table~\ref{tb:perbol58_maps} also lists the upper-level energy of each transition along with the effective density, which is the density at which the integrated intensity of the molecular line reaches 1 K \kms.

\begin{deluxetable*}{lccccccc}
\tabletypesize{\scriptsize}
\tablecaption{Per-bolo 58 Maps parameters \label{tb:perbol58_maps} }
\tablewidth{460pt}
\tablehead{
\colhead{Map} & \colhead{Rest Frequency}& \colhead{E$_{up}$$^{a}$}&\colhead{n$_{eff}$(T$_k=$10)$^{b}$}& \colhead{Array configurations} & \colhead{Synthesized beam} & \colhead{PA}& \colhead{RMS}\\
&[GHz]&[K]&[cm$^{-3}$]&&&[mJy beam$^{-1}$]}
\startdata
NH$_2$D$(1_{1,1}-1_{0,1})$ &85.926263&20.7&-& D+E & 6".9 x 5".9&75$^{\circ}$ & 65          \\
N$_2$H$^+(1-0)$& 93.17340&4.5& $1.5\times10^4$    & D+E& 6".3 x 5".3&81$^{\circ}$  &  70        \\
HCN$(1-0)$ & 88.631846&4.3&  $8.4\times 10^3$    &D+E & 6".6 x 5".6&79$^{\circ}$  & 58       \\
HCO$^+(1-0)$& 89.18853&4.3&  $9.5\times10^2$  &D+E&  6".6 x 5".6 &79$^{\circ}$  & 64     \\
CS$(2-1)$ &97.98095 &7.1& $2.3\times10^4$    &D+E & 6".1 x 5".1&72$^{\circ}$ &    79       \\
3mm continuum &90.58956&&& D+E & 6".6 x 5".6&70$^{\circ}$ & 0.5   \\

 \enddata

\tablecomments{The RMS for the molecular lines is measured using channels that are 0.1 \kms\ wide.}
\tablenotetext{a}{Upper-level energy of the transition.}
\tablenotetext{b}{Effective excitation density of the transition from \cite{2015Shirley}. This is the density which results in a molecular line with an integrated intensity of 1 K \kms, assuming a column density of the observed molecule and gas kinetic temperature of $\log{N}=14$ and 10 K, respectively. n$_{eff}$ for NH$_2$D has not been reported in the literature. For reference, the critical density for this transition is $7.5\times10^6$ cm$^{-3}$ \citep{2006MachinCollisional}. Critical densities are typically 1-2 magnitudes higher than effective densities.}

\end{deluxetable*}

\section{Results} \label{sec:results}

\begin{figure*}
\includegraphics[width=1\textwidth]{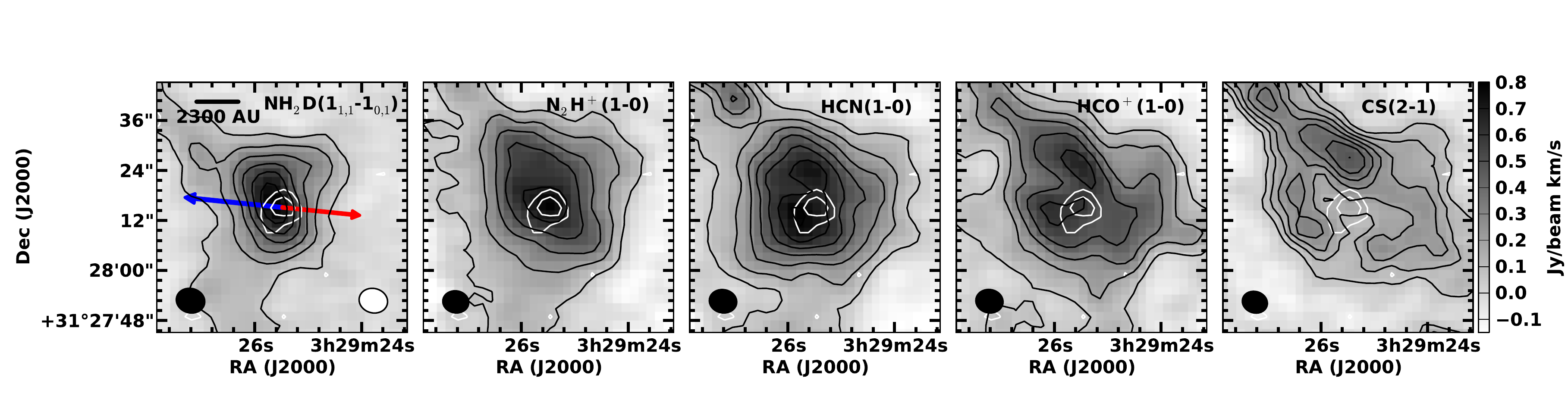}
\caption{Integrated intensity maps of different species toward Per-Bolo 58. The integration velocity range is  [7,8.1] \kms. These velocity range was chosen to include the main spectral components among the hyperfine line components. Black contours start at $3\sigma$ and increase in steps of $3\sigma$ for \cs\ and $7\sigma$ for the rest. White contours show the 3mm continuum at 2 and 3 $\sigma$ (see Table~\ref{tb:perbol58_maps}). The blue and red arrows show the direction and extent of the blue and red lobe outflow emission in \cite{2011DunhamDetection}. \label{fig:perbol58mom0}}
\end{figure*}

\begin{figure*}
   \center
   \includegraphics[width=.8\linewidth]{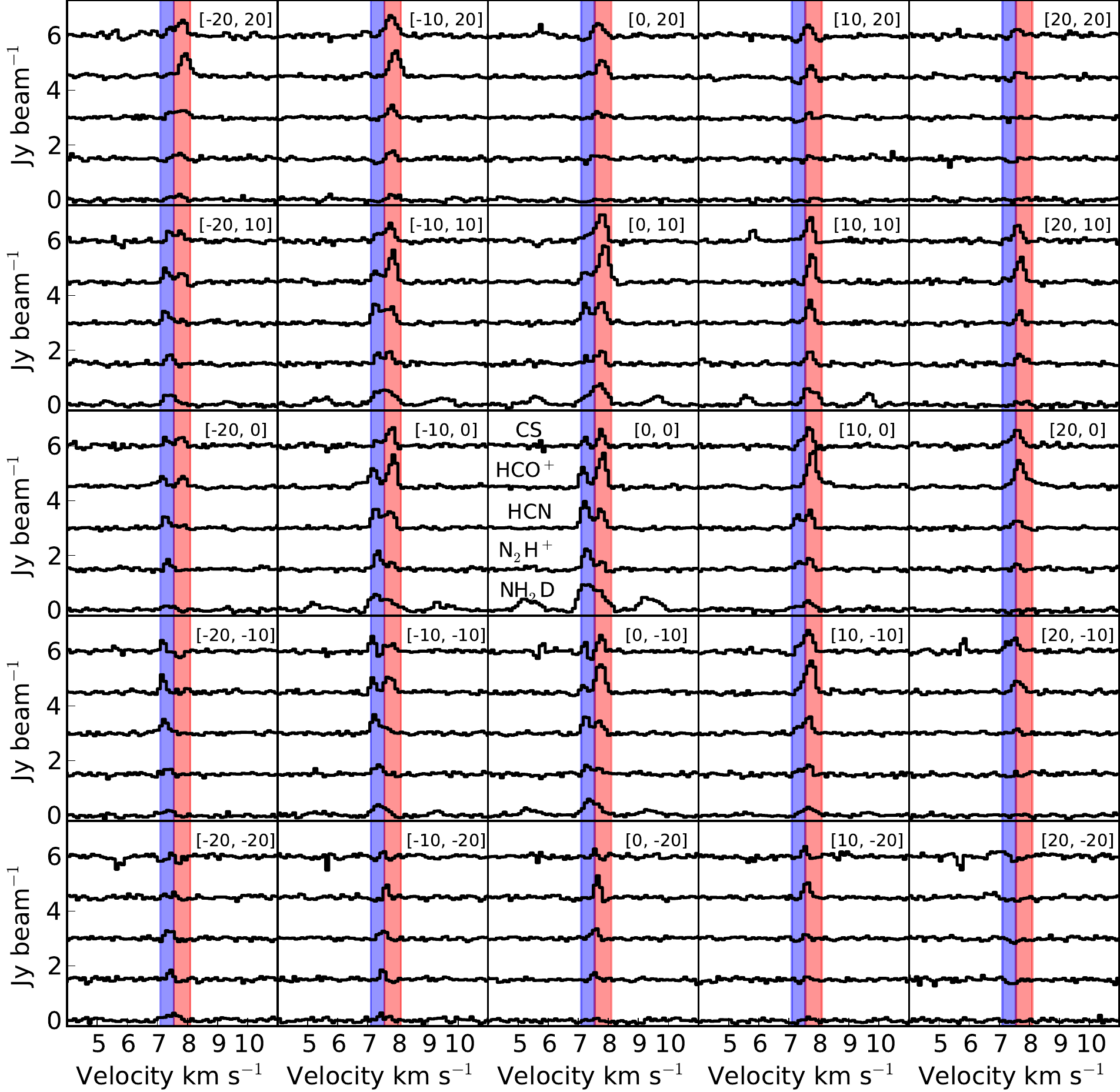}
   \caption{Molecular line spectra at different offsets (in arcseconds) from the continuum peak position. The spectra are averaged over a beam. For the \nhtd\ only the central three hyperfine lines are shown, for the \nthp\ and \hcn\ the line that is shown corresponds to the isolated and weakest hyperfine line, respectively, and therefore these hyperfine are shifted in velocity in this figure. The blue and red regions correspond to the velocity ranges used for the blue and red integrated intensity maps shown in Figure~\ref{fig:mom0percomp}.
   \label{fig:specmap}}
\end{figure*}

\begin{figure*}[t]
\center
\includegraphics[width=0.8\textwidth]{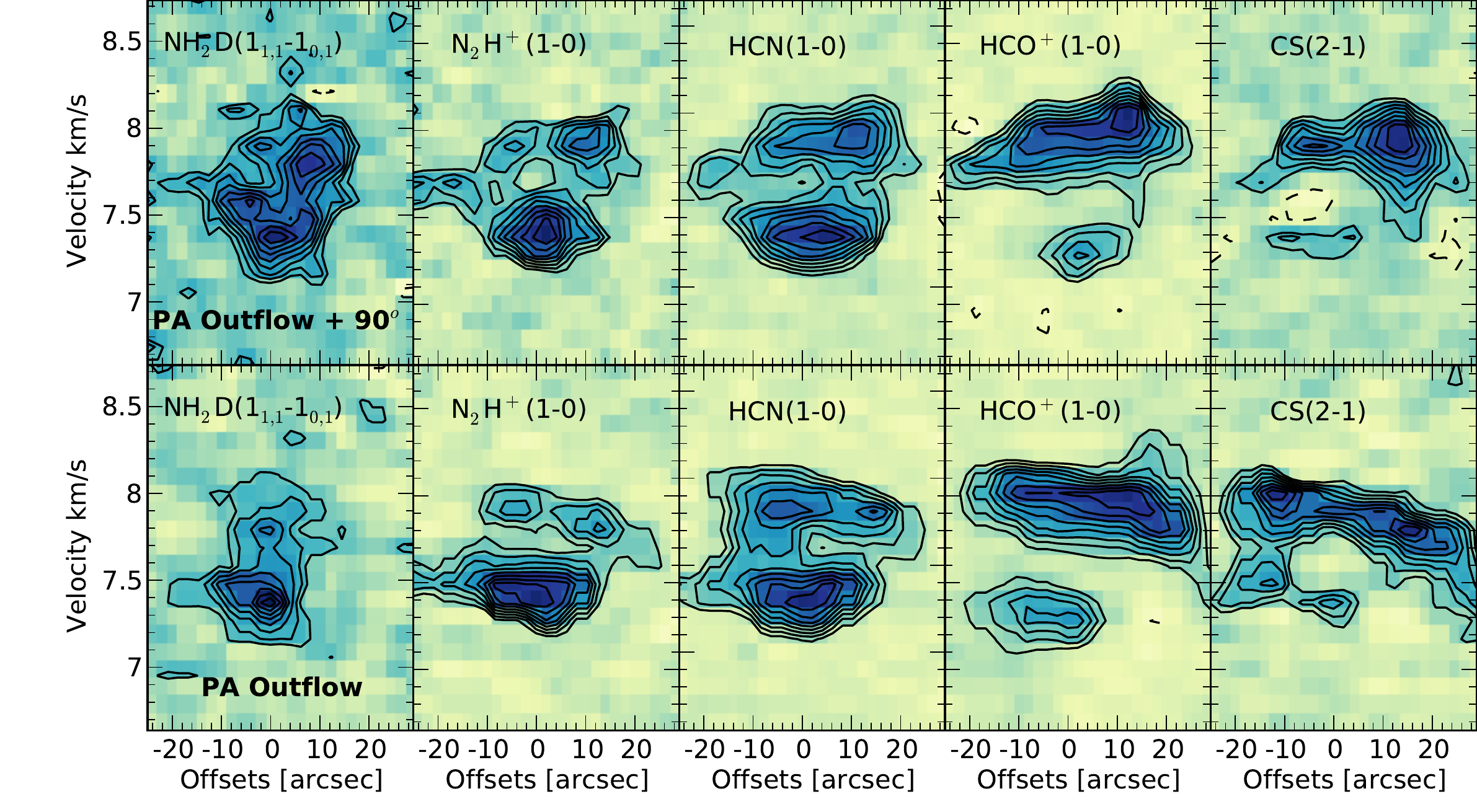}
\caption{Position-velocity maps for the molecular lines in our sample. The top and bottom row correspond to a cut in a direction perpendicular and parallel to the outflow axis, respectively. Black contours were drawn from 20 to 90 \% (in steps of 10\%) of the maximum intensity value in each panel. Black dashed contours mark regions of $-3\sigma$ values.
\label{fig:perbol58pvmap}}
\end{figure*}

Figure~\ref{fig:perbol58mom0} shows Per-Bolo 58 integrated intensity maps for all the molecules with detected emission along with the 3mm continuum in white contours. The maps show a slightly elongated structure with the long axis at a position angle of  $\sim30^{\circ}$, this is approximately perpendicular to the outflow direction. In all the observed species, except for \cs, the continuum emission lies within the contour representing 80\% of the peak emission of the integrated intensity map. The \nhtd\ and \nthp\ show a single peak which coincides with the position of the continuum source. The \hcn\ and \hcop\ show two local maxima with the continuum peak position in between them. All the molecules show extended emission toward the North-East. Extended emission directly south of the continuum source is observed in all molecular line maps, except for \cs\ where the emission extends towards the South-West. These emission tails are consistent with the large scale filamentary emission observed at 250 $\mu$m with Herschel (Figure~\ref{fig:herschel}).\\

Figure~\ref{fig:specmap} shows the spectra of all the molecules, at different offsets from the continuum peak. For \nthp\ and \hcn\ only the isolated and weakest hyperfine line are shown, respectively. All species, except for \nhtd, show two clearly distinct peaks in their spectra at the continuum peak position, separated by approximately 0.4-0.6 \kms. These two velocity components can also be seen in Figure~\ref{fig:perbol58pvmap} which shows position-velocity maps corresponding to a cut parallel and perpendicular to the outflow lobes direction. We name these two different peaks in the spectra the blue and red velocity components.

In most species, these two velocity components are also observed at positions offset from the center. The \nhtd\ line is not symmetric (including the hyperfine lines) and appears to be formed by the blending of the blue and red velocity components seen in the \nthp. Moreover, the peaks of the isolated hyperfine component of \nthp\ (typically an optically thin tracer) are not located at the velocity of the trough seen in the more optically thick lines. This is true even at several locations away from the center where instead only one component, blue or red, is present (e.g. positions at RA offset -20" for blue component and at [20",10"] for red component).\\

Our results indicate that the observed dual peak spectra are most likely due to two different velocity components rather than self-absorption and high optical depth.  Spectra with hyperfine structure can be used for opacity estimations, given that the intensity ratio between hyperfine components (assuming the same excitation temperature and width for all hyperfine lines) depends on the total opacity. Comparing the intensity ratios, using the more intense peak among the two velocity components in each hyperfine, results in estimates of the total line opacity of about 5, 4, and 9 for \nhtd, \nthp, and \hcn, respectively. We can do a similar analysis by fitting a single Gaussian to the double peak profile. The total line opacity estimate obtained this way are about 4, 5, and 6 for \nhtd, \nthp, and \hcn, respectively. For \nthp\ and \hcn, the opacity of the isolated and weakest hyperfine components respectively, is about 0.1 of the total line opacity (\citealt{2002AhrensSubdoppler,2009PaganiFrequency}). Hence, these hyperfine components in these two species (which are shown in Figure~\ref{fig:specmap}) are very likely optically thin or at most (for \hcn) moderately optically thick. We note that although it is possible that the profiles of some of the common optically thick tracer such as \hcop\ and \cs\ are being affected by opacity,  for the reasons stated above it is difficult to explain all double-peak profiles observed in this source as being due to high opacity or self-absorption.\\  

\begin{figure*}[t]

\includegraphics[width=0.98\linewidth]{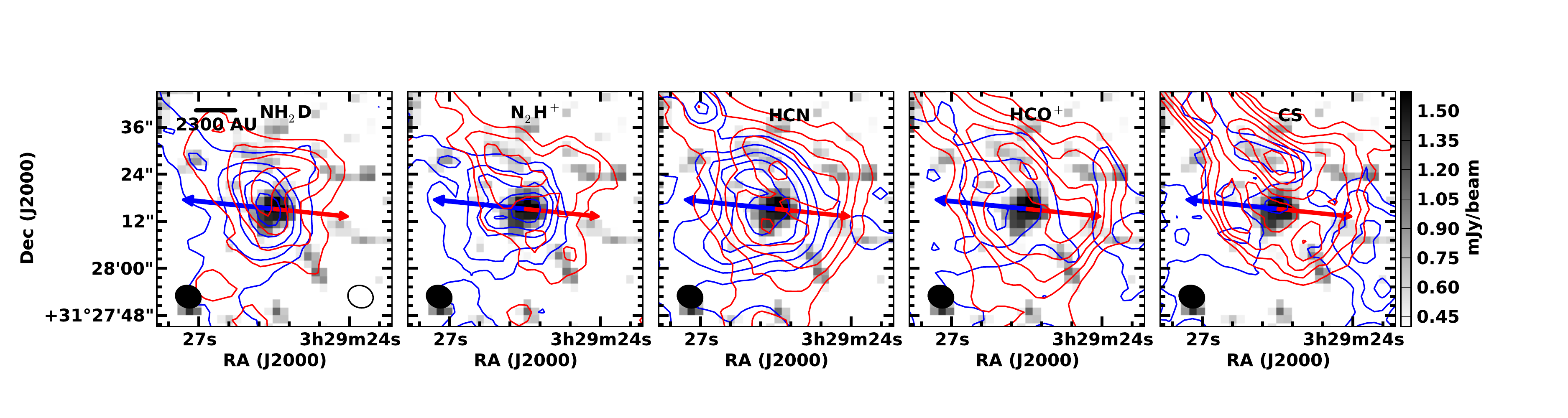}
\caption{Integrated intensity maps for the red and blue peaks in Figure~\ref{fig:specmap}. The blue (red) contours correspond to the blue (red) component. The greyscale map shows the 3mm continuum emission. The integration velocity ranges are given in the text. For \nthp, the integration is shifted to be around the isolated hyperfine component. The contours correspond to emission starting at $3\sigma$ rms and increasing in steps of $7\sigma$ for  \nhtd, \hcn\ and \hcop\ and in steps of $3\sigma$ for \nthp\ and \cs. The blue and red arrows represent the outflow as in Figure~\ref{fig:perbol58mom0}.
\label{fig:mom0percomp} }
\end{figure*}

Figure~\ref{fig:mom0percomp} shows integrated intensity maps for the blue and red velocity components shown in the spectra in Figure~\ref{fig:specmap}. The limit in velocity between the blue and red integrated intensities was selected to match the position of the trough between the two peaks at the continuum position. The lower and upper limits in velocity were selected by eye so that the integrated intensity can trace all the positions that show
emission. Using these constraints we define the blue emission as that arising from the velocity range between 7.1 and 7.5 \kms\ and the red emission as that arising from the velocity range 7.6 and 8.1 \kms. For guidance, a blue and red shaded area indicates these ranges in Figure~\ref{fig:specmap}.

Overall the blue component dominates the emission toward the East and the red component toward the West. In the cases in which the total integrated intensity (over all velocities, shown on Figure~\ref{fig:perbol58mom0}) shows two peaks (\hcn, \hcop), one corresponds to the peak in the emission of the blue component and the other to the peak emission of the red component. For \nhtd, \nthp\ and \hcn\ the blue emission peaks at and trace more closely the continuum emission. On the other hand, the \hcop\ and \cs\ emission are composed of two structures, which have less prominent peaks and at positions offset from the continuum location. Contours tracing high intensity from the red component are seen towards the North of the continuum peak in all the molecules. The relative intensity of the two components changes throughout the map, as seen in Figure~\ref{fig:specmap}. For \hcop\ and \cs\ the emission from the red component dominates at almost every position.

\section{Analysis}

The double-peaked lines of Per-Bolo 58 cannot be well reproduced by a single Gaussian fit. Thus, to quantify central velocities and linewidths we fit these spectra with two Gaussian wherever the line presented a double peak structure. To disentangle the emission coming from the blue and red components, we use the integrated intensities defined in section~\ref{sec:results} (see also Figure~\ref{fig:mom0percomp}). We fit a Gaussian to the blue (red) component if the integrated blue (red) intensity is over three times the RMS at that particular position. Thus, a single Gaussian is fit at positions were only
one integrated emission, blue or red, is over three times the RMS and two Gaussian are fit at positions where both, the blue and red integrated emission are over three times the RMS.
The free parameters in the case of molecular transitions with hyperfine structure are the central velocity $v_c$ of the main hyperfine line, the linewidth of the lines $\sigma$, the total opacity $\tau_{tot}$ and the excitation temperature $T_{ex}$. The linewidth $\sigma$ is assumed to be identical for all the hyperfine lines and $\tau_{tot}$ is the sum of all the individual hyperfine lines opacities. In the case of molecular spectra with no hyperfine structure, we fit the line with a single Gaussian with only $v_c$, $\sigma$ and the peak intensity as free parameters. For more details on the fitting functions and hyperfine frequencies used see \cite{2017MaureiraKinematics}.

\subsection{Kinematics}

Figure ~\ref{fig:perbol58allfits} shows the central velocity in columns 1 and 2 for the blue and red components, columns 3 and 4 show the ratio $\sigma_{NT}/\sigma_{T}$ between non-thermal and thermal linewidth for the two different components. The thermal contribution was calculated as $\sigma_{T}=\sqrt{k_{B}\times T/m}$ with a kinetic temperature T$=10$ K (taken from \cite{2008Rosolowsky}) and $m$ is the mass of the observed molecule. The non-thermal contribution is given by $\sigma^2_{NT}=\sigma^2-\sigma^2_{T}$, where $\sigma$ corresponds to the total linewidth obtained from the fit to the spectrum. The plot only shows pixels in which the value of the parameter obtained from the fit is at least three times its error. Maps for \nhtd\ are not shown since the blue and red peak in this molecule were blended, making the velocity and linewidth of this molecule velocity components highly uncertain.

\subsubsection{Central velocities}

Figure~\ref{fig:perbol58allfits} shows that the velocity fields of the velocity components are complex. We identify one observable trend across our molecular line sample. The blue component shows higher blue-shifted velocities at positions close to the continuum location. High red-shifted velocities can also be seen close to the continuum location for the red component, although the highest red-shifted velocities are seen towards the North-East tail. This trend towards the center is more clearly seen in the \nthp\ velocity maps, as well as in the p-v diagram of this molecule and the \nhtd\ in Figure~\ref{fig:perbol58pvmap}.

We note that if we fit linear velocity gradients to maps produced by a single Gaussian profile to each spectrum (i.e., including those with double peak profile) we find gradients along the minor axis of the elongated core structure shown in Figure~\ref{fig:perbol58mom0} for the \nhtd, \nthp\ and \hcn. On the other hand, since the \hcop\ and \cs\ spectra are dominated by the red velocity component, they show gradients along the major axis of the elongated structure, similar to Figure~\ref{fig:perbol58allfits}.

\subsubsection{Linewidths}

The linewidths in Figure~\ref{fig:perbol58allfits} also show a complex field for both components, where most of the molecular lines show several peaks. All molecular lines have non-thermal linewidths that are less than the H$_2$ molecule thermal sound speed at $T=10$ K, except in some regions in the \hcop\ and \cs\ lines, where the non-thermal contribution is close to the sound speed (East of the continuum peak and North-East tail). Interestingly, extended regions with narrower linewidths are located preferentially on the non-overlapping (or outskirts) of both components for \nthp\ and \hcn. The non-thermal linewidths among the red and blue components are similar for \nthp, while larger for the red component in the \hcn. Similarly, larger non-thermal linewidths are found in the red rather than the blue component for \hcop\ and \cs.

We note that if both velocity components were fit by a single Gaussian, a broad linewidth region would be seen at the intersection of both velocity components for \nhtd\ (Figure~\ref{fig:perbol58pvmap}) and \nthp\ and then, a trend of decreasing linewidth with distance from the continuum position. The opposite would be observed for \hcop\ and \cs, the largest linewidths are typically located on the outskirts and are traced mainly by the red velocity component. The latter is produced by a combination of broad linewidths regions located at the North-East tail and at locations where the outflow might be impacting the envelope (see below).

\subsubsection{Outflow-envelope interaction}

\begin{figure*}
\includegraphics[width=1\textwidth]{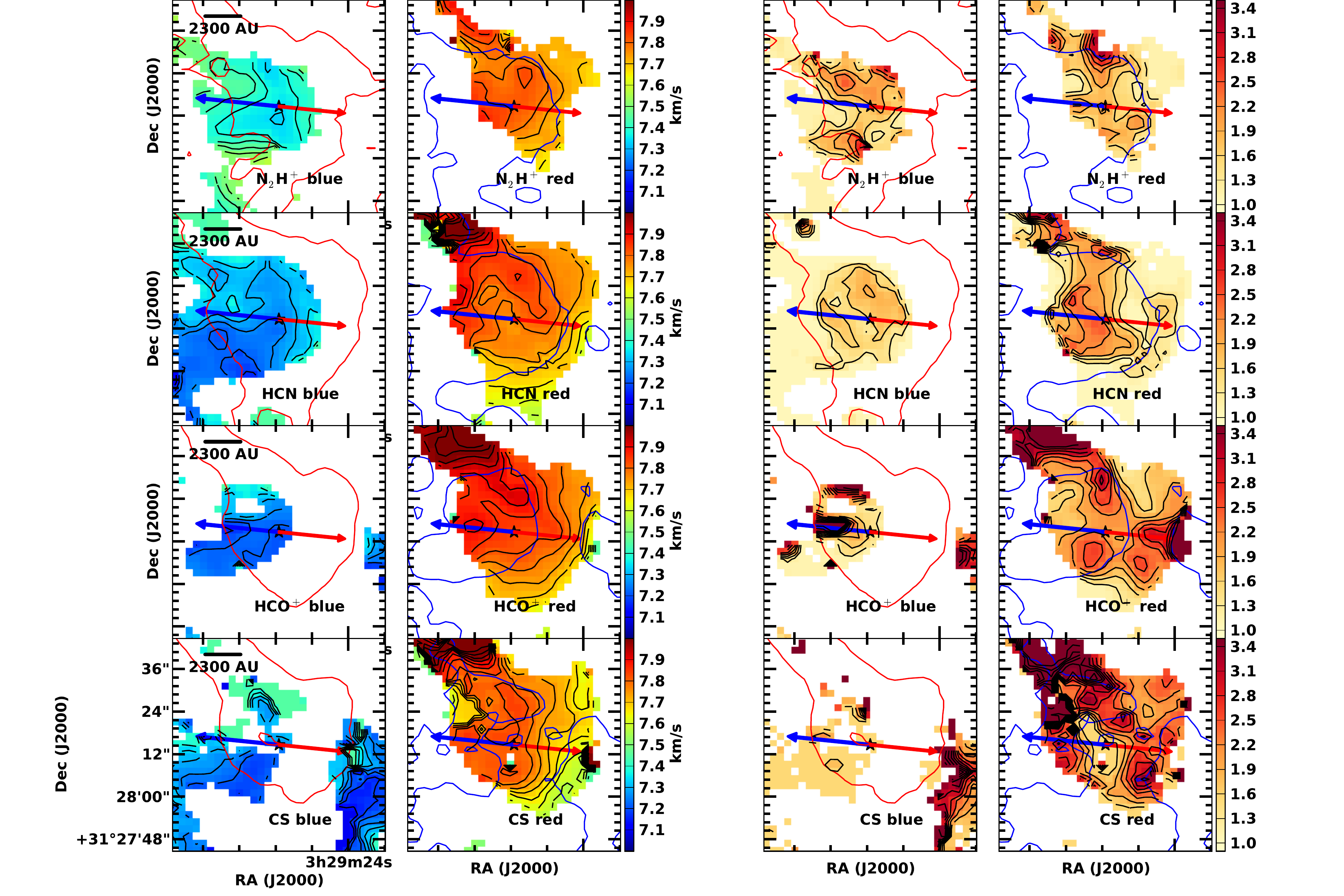}
\caption{Central velocity maps for the blue (column 1) and red (column 2) components using the spectral maps of \nthp, \hcn, \hcop\ and \cs. Maps of the ratio of non-thermal to thermal (at 10 K) linewidths for the blue and red components are shown in columns 3 and 4, respectively. For guidance, red and blue contours trace the integrated intensity of the red and blue components at 10$\%$ (\nthp, \hcn) and 30$\%$ (\hcop, \cs) of the maximum integrated intensity. Black contours follow the velocity from 6.9 to 8.1 \kms, in steps of 0.04 \kms for columns 1 and 2 while they follow the linewidth ratio from 1 and increasing in steps of 0.3 for columns 3 and 4. The black star marks the position of the continuum and the blue and red arrows represent the outflow as in Figure~\ref{fig:perbol58mom0}.
\label{fig:perbol58allfits}}
\end{figure*}

The \hcop\ shows evidence of outflow-envelope interactions in the central velocity and linewidth distribution. Some regions of high redshifted and blueshifted velocities are consistent with the direction of the outflow lobes. Similarly, relatively large linewidths ($\gtrsim$0.17 \kms) are observed along the direction of the outflow lobes (see Figure~\ref{fig:perbol58allfits}), where low-intensity wings towards high-velocity blue and red velocities, east, and west of the continuum source, respectively, were detected (see Figure~\ref{fig:specmap} and~\ref{fig:perbol58pvmap} for the high-velocity red wing).

\subsection{Masses}
\label{sec:massesandstability}

We calculate the mass of each velocity component from the N$_2$H$^+$ emission assuming local thermodynamics equilibrium (LTE) following the procedure in section 4.1 of \cite{2017MaureiraKinematics}. We use a fractional abundance for N$_2$H$^+$ of $1.5\cdot10^{-9}$. This value is consistent with the value reported by \cite{2013DanielNitrogen} for the B1b core which is also in Perseus and contains the FHSC candidates B1b-N and B1b-S \citep{2012PezzutoHerschel,2013HuangProbing,2014HiranoTwo}. Table~\ref{tb:perbol58ltemmasses} lists the resultant LTE masses, as well as the peak column density and the total N$_2$H$^+$ mass for the blue and red components. These are calculated using the emission within half of the FWHM of a 2D Gaussian fit to the integrated intensity for each component, which is about 4100 au for both components, centered on the peak of the 2D Gaussian fit (that is, we do not include in our mass estimate the emission extending in a tail-like structure towards the northeast and the emission extending towards the south seen in many of the molecular line maps). Both the blue and the red components have a similar mass of $\sim0.5-0.6$ M$_{\odot}$. The total mass of the sum of the two components ($\sim$1.1 M$_{\odot}$) is within the range 0.8-2.7 M$_{\odot}$ given by envelope mass estimates from dust continuum observations at 1.1mm and 3mm \citep{,2006EnochComplete,2007HatchellStarb,2010EnochCandidate,2010SchneeObserved}. \\

We can compare, approximately, the gravitational potential energy and kinetic energy of the structures that give rise to the two different velocity components in the spectra to test if they are a gravitationally bound system and thus, likely to be infalling as part of Per-Bolo 58 envelope. For calculating the kinetic energy of a component we assume that the difference in velocity between the two components along the line of sight $\Delta v_{los}\sim0.4$ \kms\ is representative of the difference between velocities in the other directions. This difference in velocity is larger than the sound speed and the local non-thermal velocity dispersion. The total velocity difference between the two components is then $\sqrt{3}\Delta v_{los}$ and the kinetic energy of a component of mass $m_{i}$ is calculated as $\frac{1}{2}m_{i}(\sqrt{3}\Delta v_{los}/2)^2$. Similarly, for the potential energy calculation, we assume that the separation in the plane of the sky $r_{pos}\sim2300$ au is a good approximation for the separation along the line of sight. Then, the total separation would be given by $\sqrt{2}r_{pos}$ \citep{2015PinedaFormation}, and the potential energy of the pair is given by $-\frac{Gm_{1}m_{2}}{\sqrt{2}r_{pos}}$. The ratio of the kinetic to the potential energy for the red and blue component system is then $\sim0.8$. Hence, these two velocity components are consistent with being gravitationally bound at the envelope scales of few thousands au.

\begin{deluxetable}{cccrrrrrrrcrl}
\tabletypesize{\scriptsize}
\tablecaption{Per-Bolo 58 \nthp\ LTE Mass estimates \label{tb:perbol58ltemmasses} }
\tablehead{
\colhead{Component}&\colhead{N$^{peak}$(N$_2$H$^+$)} & \colhead{Mass(N$_2$H$^+$)} & \colhead{Total (H$_2$) Mass}  \\
&\colhead{[cm$^{-2}$]}&\colhead{M$_{\odot}$}&\colhead{M$_{\odot}$}}
\startdata

Blue &   $7-10(\times10^{12})$  &$1.1-1.3(\times10^{-9}$) &$0.5-0.7$\\
Red &  $7-10(\times10^{12})$&$0.9-1.1(\times10^{-9}$)& $0.4-0.6$   \\

\enddata

\tablecomments{ The lower and upper limits in each case correspond to the estimate made with Gaussian fits that assumed a fixed excitation temperature of 10 K and 6 K, respectively. These masses are calculated within a radius of $4100$ au of the center of the blue and red integrated intensity. We use a fractional abundance for the N$_2$H$^+$ of $2\cdot10^{-9}$.  A distance of 230 pc was assumed.}

\end{deluxetable}

\section{Origin of the two velocity components}
\label{sec:turbulentorigin}

In this section we discuss two possible scenarios for the origin of the two velocity components in the envelope of Per-bolo 58. In the first scenario the two velocity components are produced by infall motions on two sides of an inhomogeneous and flattened envelope. In the second scenario, the two velocity components are a consequence of accretion from large scale filamentary structures at different velocities.

\subsection{Turbulence at the core scale}

Given the morphology and velocity structure of the emission associated with the two components, one interpretation is that they are the product of infall in a flattened and inhomogeneous envelope. If we were to view a system closer to edge-on rather than pole-on, then the infall in an envelope with some degree of flattening would appeared as a velocity gradient where the blue-shifted velocities appear on the side of the blue outflow lobe and the red-shifted velocities on the side of the red outflow lobe, as seen in our molecular line maps of Per-Bolo 58. If the distribution of the gas in this envelope is inhomogeneous, then the overlapping region between the red-shifted side and the blue-shifted side of the envelope can be detected as two optically thin velocity components, instead of one optically thick emission line with two peaks as expected in a more symmetric collapse \citep{2011TomisakaObservational}. This is what it appears we are seeing in Per-Bolo 58. Such envelope structure can arise from the collapse of a magnetized, initially perturbed core. Turbulence affects the structure of the envelope by making the gas more clumpy and filamentary \citep{2008OffnerDriven,2011SmithQuantification,2015SeifriedAccretion}. Magnetic fields produce the flattening of the envelope with time since they give a preferential direction for the infall along the magnetic field lines.\\

In order to test this hypothesis, we analyze a magneto-hydrodynamic (MHD) simulation of an isolated 4 $\msun$ collapsing core where the initial conditions include an ordered magnetic field and a turbulent velocity field. This core mass is close to the estimated range for Per-Bolo 58 of 1-3 $M_{\odot}$ (\citealt{2006EnochComplete,2007HatchellStarb,2010EnochCandidate,2010SchneeObserved}) and thus the simulation can be used to compare general kinematic properties.

\subsubsection{Description of the simulation}

The simulation begins with a uniform density sphere of 10 K gas with a radius of 0.065 pc and mass 4 M$_{\odot}$ ($\rho=2.4\cdot10^{-19}$ g cm$^{-3}$). The core is threaded by a magnetic field in the z direction with a normalized mass-to-flux ratio of 2.5. At $t=0$, the gas velocities in the core are perturbed with a turbulent random field that has a flat power spectrum over wavenumbers $k=$1-2. The initial gas velocity dispersion is 0.52 \kms, which corresponds to a viral parameter of 2. Once set in motion, the initial turbulence decays and no additional energy injection occurs.

We note that in our observations the N$_2$H$^+$ does not seem to be affected by freeze-out or outflow motions, unlike the more optically thick tracers in our sample, such as HCO$^+$ and CS. Given that our simulation does not include outflow motions or freeze-out chemistry, we only use the N$_2$H$^+$ to compare with the observations. We produce \nthp\ maps for several different views at times $\sim$0.13-0.16 Myr. These times coincide with the main collapse phase and extend until just after the formation of a protostar.  We use the radiative transfer code {\sc radmc-3d}\footnote{\url{http://www.ita.uni-heidelberg.de/~dullemond/software/radmc-3d/}} to calculate the \nthp\ emission. We use the excitation and collisional data from the Leiden atomic and molecular database and include the hyperfine structure \citep{schoier05}. We perform the radiative transfer using the non-LTE Large Velocity Gradient approximation \citep{shetty11}.  We first flatten the adaptive mesh refinement data to a fixed 256$^3$ resolution for regions with sizes of 0.065 pc ($\Delta x=$52 au). To convert the simulation mass densities into N$_2$H$^+$ densities, we adopt an abundance of $10^{-10}$ N$_2$H$^+$ per H$_2$. This value is lower than what we expect for Per-Bolo 58 given the comparison between the \nthp\ mass in section~\ref{sec:massesandstability} and dust mass from \cite{2010EnochCandidate} and \cite{2010SchneeObserved}, however, this does not affect our comparison of the kinematics and morphology of Per-Bolo 58 and the simulation. We set the abundance to zero for gas with temperatures above 800 K to exclude the warm ambient medium surrounding the core. To account for small-scale turbulence below the simulation grid scale, we include a turbulent broadening of 0.02 kms$^{-1}$, which is obtained from the linewidth-size relation at the grid resolution \citep{2007McKeeTheory}. Each emission cube was centered at 93.176 GHz. The velocity channels span $\pm 12$ kms$^{-1}$ and have a width of 0.09 kms$^{-1}$. More details of the simulations can be found in \cite{2016DunhamALMA}, \cite{2016OffnerTurbulent} and \cite{2017OffnerImpact}.

\subsubsection{Overview of the Collapse in the Simulation}

\begin{figure*}
\centering
\includegraphics[width=0.8\textwidth]{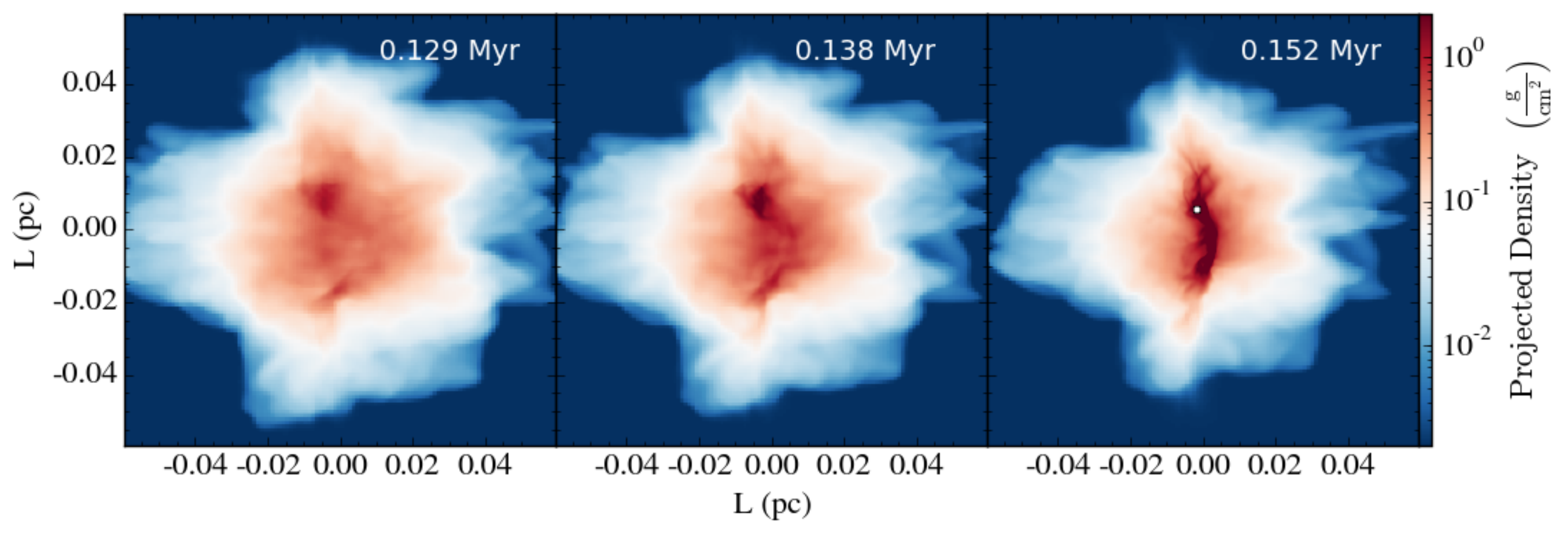}
\caption{Simulation column density maps for the core at 3 different times, viewed edge-on. The sink particle location, after it forms, is marked on the right-most panel. The magnetic field is mainly along the horizontal direction.\\
\label{fig:simulationdensity}}
\end{figure*}

Here we will briefly describe the collapse in the simulations at $\sim1000$ au scales. As collapse begins, turbulence creates slight over densities in the dense gas and some mixing occurs between the core and low-density gas. Over time, the core becomes asymmetric and centrally condensed. At around $\sim$0.13 Myr it becomes apparent that gas moving towards the center is doing so mainly in a direction aligned with the magnetic field and that the gas at the center is contracting also preferentially along the magnetic field. Figure~\ref{fig:simulationdensity} shows column density maps for the core viewed edge-on, at different time steps. The initial magnetic field in these figures is along the horizontal direction. At $\sim$0.15 Myr (Figure~\ref{fig:simulationdensity} right panel) the dense structure formed at the center is highly flattened along the magnetic field direction, such that it looks filamentary when viewed perpendicular to the magnetic field. The envelope of infalling gas surrounding this structure shows an irregular morphology.

\subsubsection{Synthetic Observations}

We analyze the simulation data cube at two different time steps, 0.13 Myr and 0.15 Myr, and at two different inclinations, perpendicular to the magnetic field (edge-on) and along the magnetic field (pole-on). The 0.13 Myr time step corresponds to a prestellar stage, while the 0.15 Myr output models the gas distribution expected just after the time of first core formation. The sink particle in the later case has a mass of 0.016 $\msun$. We first study the spectra of these cubes to see if they show two velocity components along the line of sight. Two velocity components along the line of sight are visible for the two different time steps and for the two different views. In order to make the comparison with the observations we first re-sampled the velocity channels of the simulation cubes to match the observation resolution, and shifted the spectrum at the center of the map so the main component of the \nthp\ spectrum is at the same velocity as the main component in Per-Bolo 58 at the continuum peak position. We also matched the position of the Per-Bolo 58 continuum source with the coordinates of the center of the core in the simulation and assumed that the simulated core is observed at the same distance as Per-Bolo 58 ($\sim$ 230 pc). Then, using the MIRIAD task uvmodel, we generated visibilities for the simulated cubes using the uv-sampling of the observed Per-Bolo 58 \nthp\ data. Finally, we generated the synthetic cleaned cubes by applying the same imaging procedures as for the observed \nthp\ cube. The two velocity components remained detectable in the synthetic cubes.

\subsubsection{Comparison with observations} \label{sec:comparison}

\begin{figure}
\includegraphics[width=0.47\textwidth]{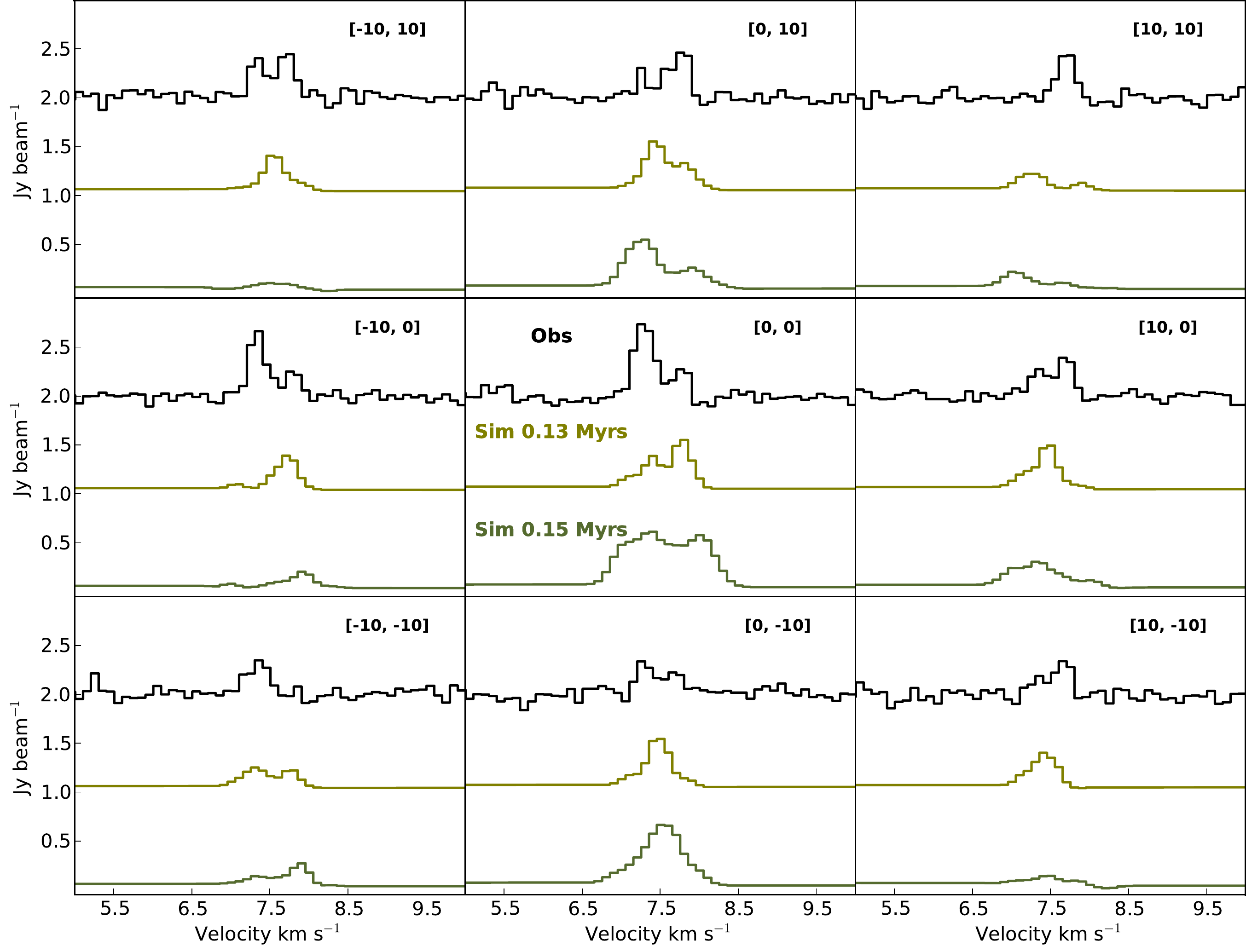}
\caption{\nthp\ spectra at different offsets (in arcseconds) from the continuum peak position (Per-Bolo 58, black) and center position (simulations, green). The simulated spectra correspond to the edge-on view data cube in Figure~\ref{fig:simulationdensity} where the light and dark green color correspond to the time steps 0.13 and 0.15 Myr, respectively. \label{fig:perbol58simspecmap}}
\end{figure}

\begin{figure*}
\includegraphics[width=1\textwidth]{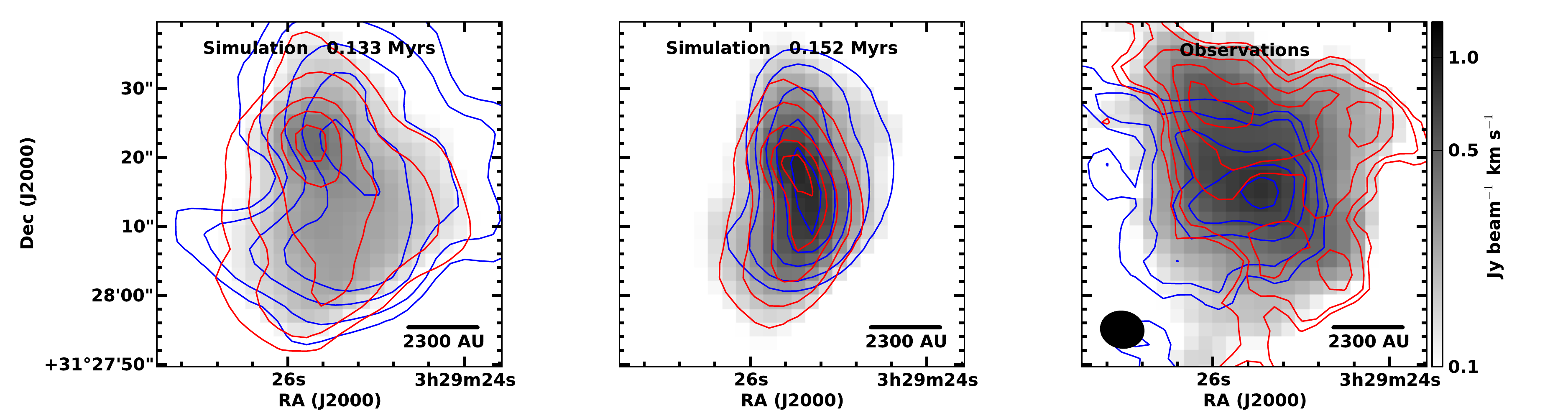}
\caption{Simulations and Per-Bolo 58 integrated intensity maps for the blue and red peak in the spectra of \nthp. The simulation corresponds to synthetic observations made with the same uv-sampling as the observed source. The view of the simulations is edge-on. The magnetic field and the net angular momentum vectors in the simulations are along the horizontal direction. The gray maps show the total (blue plus red) integrated intensity in each map. The colorbar is in units of Jy beam$^{-1}$\kms\ and it is presented in a logarithmic scale. The blue(red) contours show the blue(red) component integrated intensity. Left: simulation at 0.133 Myr.  Middle: simulation at 0.152 Myr. Right: observations. Black ellipse corresponds to the beam. Contours are drawn at 20, 30, 50, 70 and 90$\%$ of the maximum for the observations and simulations.
\label{fig:perbol58simmom0}}
\end{figure*}

Figure~\ref{fig:perbol58simspecmap} shows \nthp\ spectra at different positions and at two different time steps in the simulation viewed edge-on along with the spectra from the CARMA observations of Per-Bolo 58. The two different time steps show two velocity components (although sometimes they are blended due to the limited resolution) in some locations. In other locations the simulation shows spectra with one dominant (blue or red) spectral component, similar to the observed \nthp\ spectral map of Per-Bolo 58.\\

Figure~\ref{fig:perbol58simmom0} shows the \nthp\ integrated intensity map for the blue and red velocity components for these same simulation outputs and for Per-Bolo 58. We selected the ranges for integration following the same procedure described for the observations (see Section 3). Similar to the observations, the two velocity components show spatial overlap. The blue and red integrated intensity peaks are close to each other in both of the simulation time steps in Figure~\ref{fig:perbol58simmom0}, resembling most of the emission distribution seen for both components in most of the observed molecular line maps (see Figure~\ref{fig:mom0percomp}). The more compact contours in the simulation follow the elongated dense structure that forms during the collapse. This structure is perpendicular to the direction of the magnetic field. The later stage shows a more compact integrated emission in both components in comparison with both our observations and the earlier simulation output. This is a result of the evolution of the denser regions in the envelope, which become more flattened with time. Thus, morphologically Per-Bolo 58 \nthp\ emission agrees better with the earlier evolutionary stages in the simulations (i.e., $t\sim0.13$ Myr) and/or with a system viewed close to, but not exactly, edge-on. \cite{2011DunhamDetection} concluded using the outflow morphology that extreme edge-on or pole-on inclinations are ruled out, and thus, an inclination different from exactly edge-on is expected. Given the close spatial overlap between the blue and red components (see Figure~\ref{fig:perbol58simmom0}), inclinations closer to edge-on rather than pole-on are in better agreement with Per-Bolo 58. Indeed, the same simulation viewed pole-on shows little spatial overlap between the components, as expected if the emission arises from a flattened infalling envelope. Similarly, as mentioned in Section~\ref{sec:turbulentorigin}, a velocity gradient where the blue-shifted velocities appear on the side of the blue outflow lobe and the red-shifted velocities on the side of the red outflow lobe is also consistent with a system that is close to edge-on. \\

\begin{figure*}
\center
\includegraphics[width=0.8\textwidth]{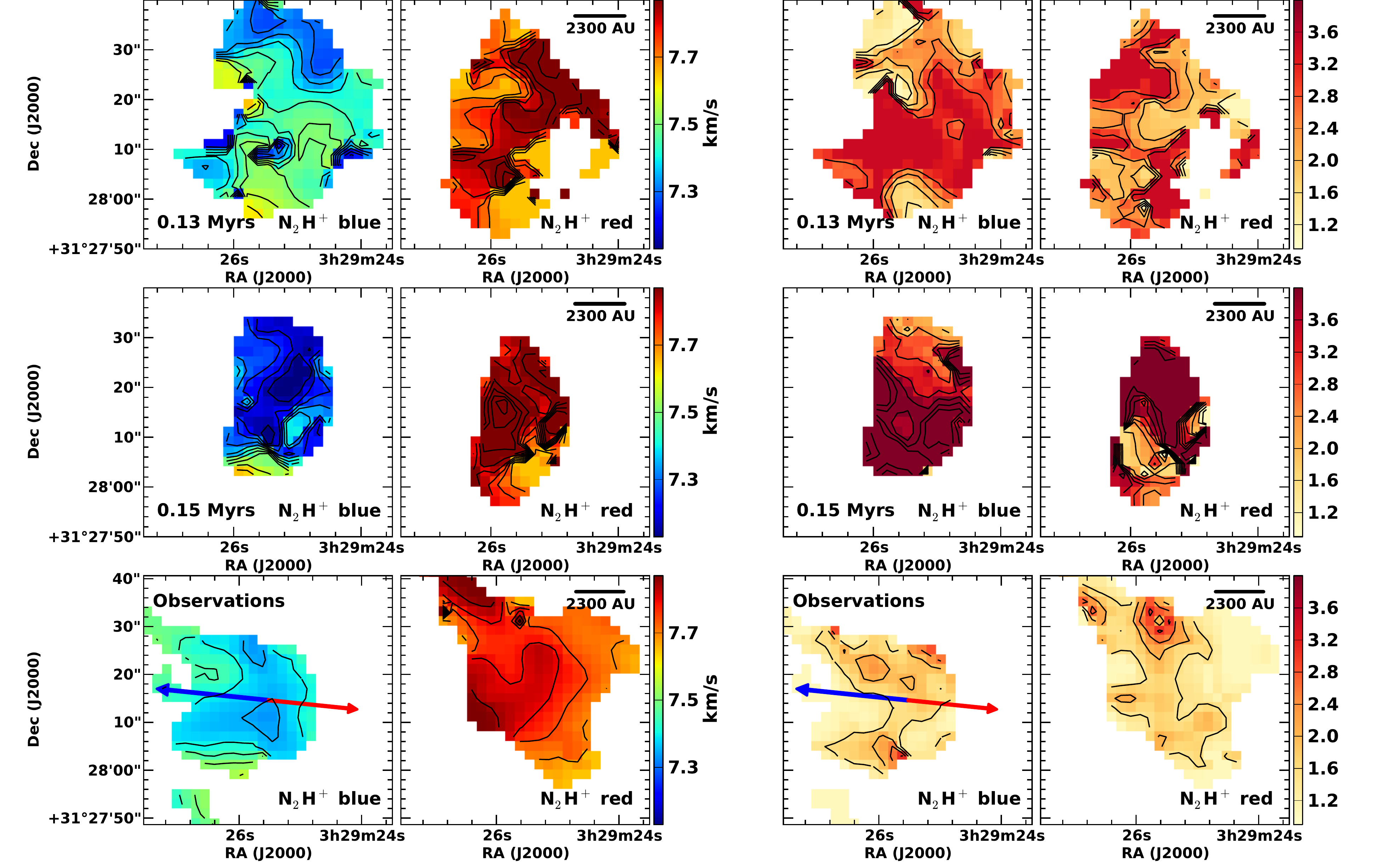}
\caption{Simulation and Per-Bolo 58 central velocity (left) and linewidth (right) maps for the blue and red peak in the spectra of \nthp. Values in these maps were obtained from a double Gaussian fit. Contours for the velocity maps are drawn from 7.1 \kms, increasing in steps of 0.05 \kms. Contours for the non-thermal over thermal linewidth ratio maps are drawn from 1, increasing in steps of 0.5. The blue and red arrows represent the outflow as in Figure~\ref{fig:perbol58mom0}. \label{fig:perbol58simallfits}}
\end{figure*}

\begin{figure}
\center
\includegraphics[width=0.5\textwidth]{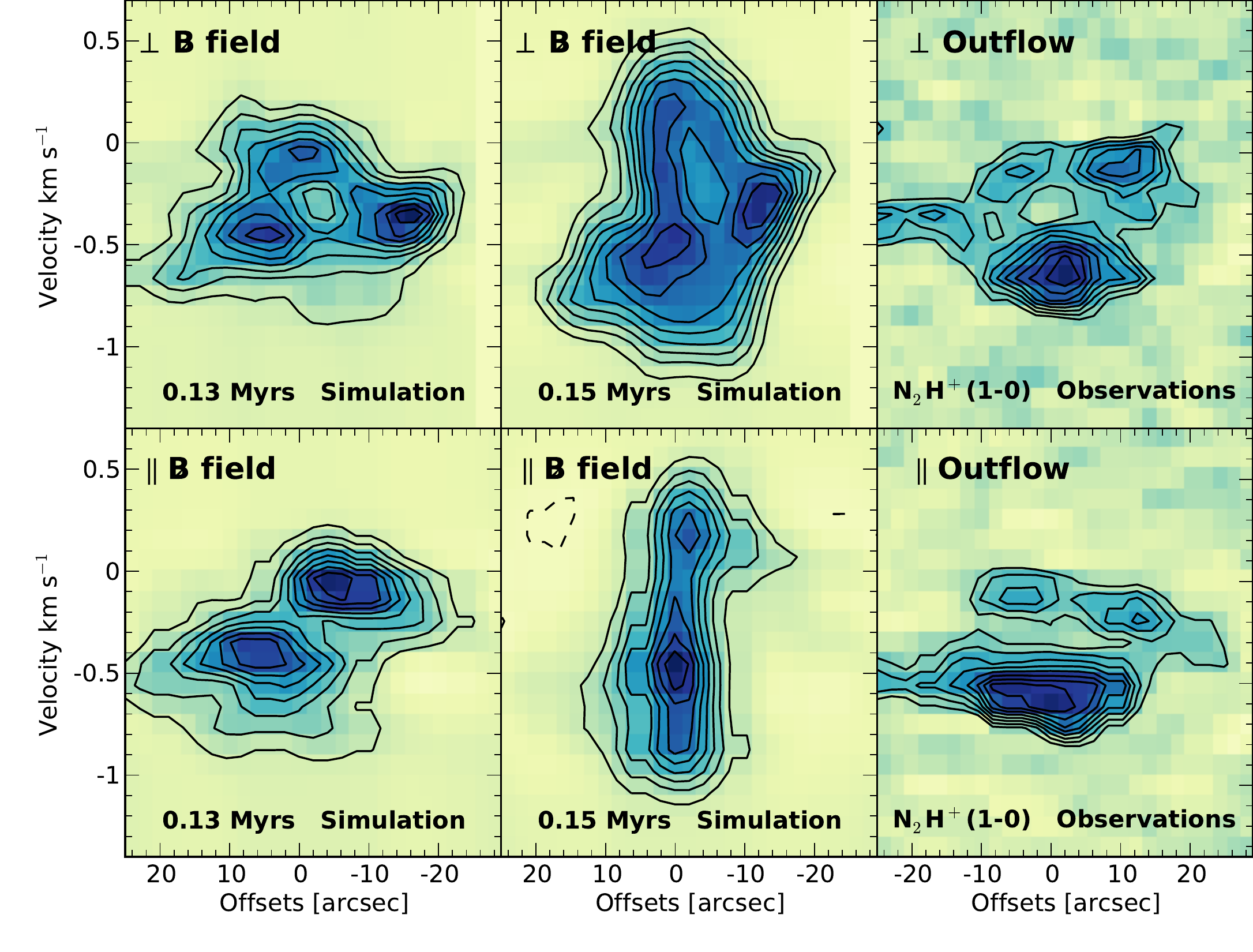}
\caption{\nthp\ position velocity maps for the simulations at 0.13 and 0.15 Myr. For comparison, Per-Bolo 58 position velocity maps for \nthp\ are also shown. For the simulations, the top and bottom panels are cut perpendicular and parallel to the B-field direction, respectively. For the observations the top and bottom row correspond to a cut in a direction perpendicular and parallel to the outflow lobes direction, respectively. Black contours were drawn from 20 to 90 \% (in steps of 10\%) of the maximum intensity value in each case. Black dashed contours mark regions of negative 3 sigma values in the observations.
\label{fig:sim_pvmap}}
\end{figure}

Figure~\ref{fig:perbol58simallfits} shows \nthp\ maps of velocity and non-thermal over thermal linewidth ratio for the view and time steps previously shown. We also show Per-Bolo 58 for comparison.

The collapse at 0.13 Myr shows a more complex and disordered velocity field than the collapse at 0.15 Myr. In comparison, the velocity field in Per-Bolo 58 shows less sub-structure than the simulated collapse at both stages. The absolute velocity difference between the blue- and red-shifted velocities increase from the early to the later simulation. This can also be seen in Figure~\ref{fig:sim_pvmap} which compares p-v diagrams of the simulated collapse at both time steps with Per-Bolo 58 observations. The later stage shows a concentration of blue-shifted and red-shifted velocities close to the center, which results from the turbulent conditions of the gas in the envelope. This velocity distribution is similar to what is seen in Figure~\ref{fig:perbol58allfits} for Per-Bolo 58. \\

A mix of both narrow and broad linewidths are present in both, the simulation and observations. However, linewidths are higher in the simulation at both stages, as compared to observations. The non-thermal contribution is up to $\sim$3 and $\sim$4 times the thermal one for the simulation at 0.13 and 0.15 Myr, respectively\footnote{The value of the non-thermal linewidth for the simulation at 0.15 Myr is 1.5 to 2 times the sounds speed at 10K.}. Although, the non-thermal contribution is also up to $\sim3$ times the thermal one for observations, the regions that show these large linewidths are not as extended in Per-Bolo 58 as in the 0.13 Myr old simulation. We note however, that the extended regions of broad linewidths in the simulations are in several cases a product of the blending of the two velocity components, in which case the double Gaussian fitting can give uncertain results. For instance, the fit can be equally good if one velocity components is fit significantly wider than the other or if the two velocity components are fit with similar narrower linewidths. Thus, the comparison between the blue and red components linewidths is somewhat uncertain in the case where the lines from the two components blend.

The linewidths of the simulations would probably agree more with those observed for Per-bolo 58 if for example the initial turbulence in the simulation had a smaller amplitude (which would be expected in a less massive core such as Per-Bolo 58) and/or the simulated core were viewed at an inclination different from exactly edge-on. Indeed, when viewed pole-on, the same simulations show smaller linewidths. This implies that asymmetric infall in the simulations contributes to the the non-thermal linewidth, which increases with inclinations approaching edge-on.

\subsubsection{Comparison with other sources}

A model with a flattened and infalling envelope has also been proposed for the prestellar source L1544, for which molecular line profiles similar to Per-bolo 58 have been reported. \cite{1998TafallaL1544} and \cite{2002CaselliMolecular} studied the prestellar core L1544 and found that \nthp\ as well as the low abundance species D$^{13}$CO$^+$(2-1), HC$^{17}$O$^+$(1-0) and C$34$S(2-1) show two velocity components. \cite{2002CaselliMolecular} compared the \nthp\ line profiles of L1544 with a contracting disk-like structure (seen close to edge-on), where the infall occurs in the mid-plane direction but without the inclusion of initial turbulence. They were able to produce double-peaked profiles only when a central hole of about 2000 au was included in the model (which can be a way to model depletion). Without this central hole, thermal broadening (at a temperature of 10 K) erases the double peak profile produced by the blue-shifted and red-shifted side of the structure at all positions except towards the center of the core. Unlike, L1544, our observations do not suggest \nthp\ depletion in Per-Bolo 58, as the peak of the \nthp\ integrated intensity coincides with the continuum location (see Figure~\ref{fig:perbol58mom0}). We were able to reproduce the double peak profile, without including depletion, due to the inhomogeneity in the gas created by the turbulence.
We note that both the preferred model to explain the two peaks in the profile of optically thin tracers in L1544 and the one discussed above for Per-bolo 58, require an irregular or discontinuous distribution of the infalling gas (rather than a contiguous and spherically symmetric one).

\subsection{Accretion from large scale filament}

Per-bolo 58 is located within a relatively large ($\sim$6' or $\sim$0.4 pc long) filamentary structure that extends in a northeast-southwest direction (see Figure~\ref{fig:herschel}). It is possible that the velocity structure seen in the envelope (at scales of $\sim$1000 au) are the result of motions at the filament scale. For instance, the core could be accreting material from the filament as seen in simulations of cores formed in turbulent molecular clouds \citep{2011SmithQuantification}. \\

\begin{figure}
\center
\includegraphics[width=0.5\textwidth]{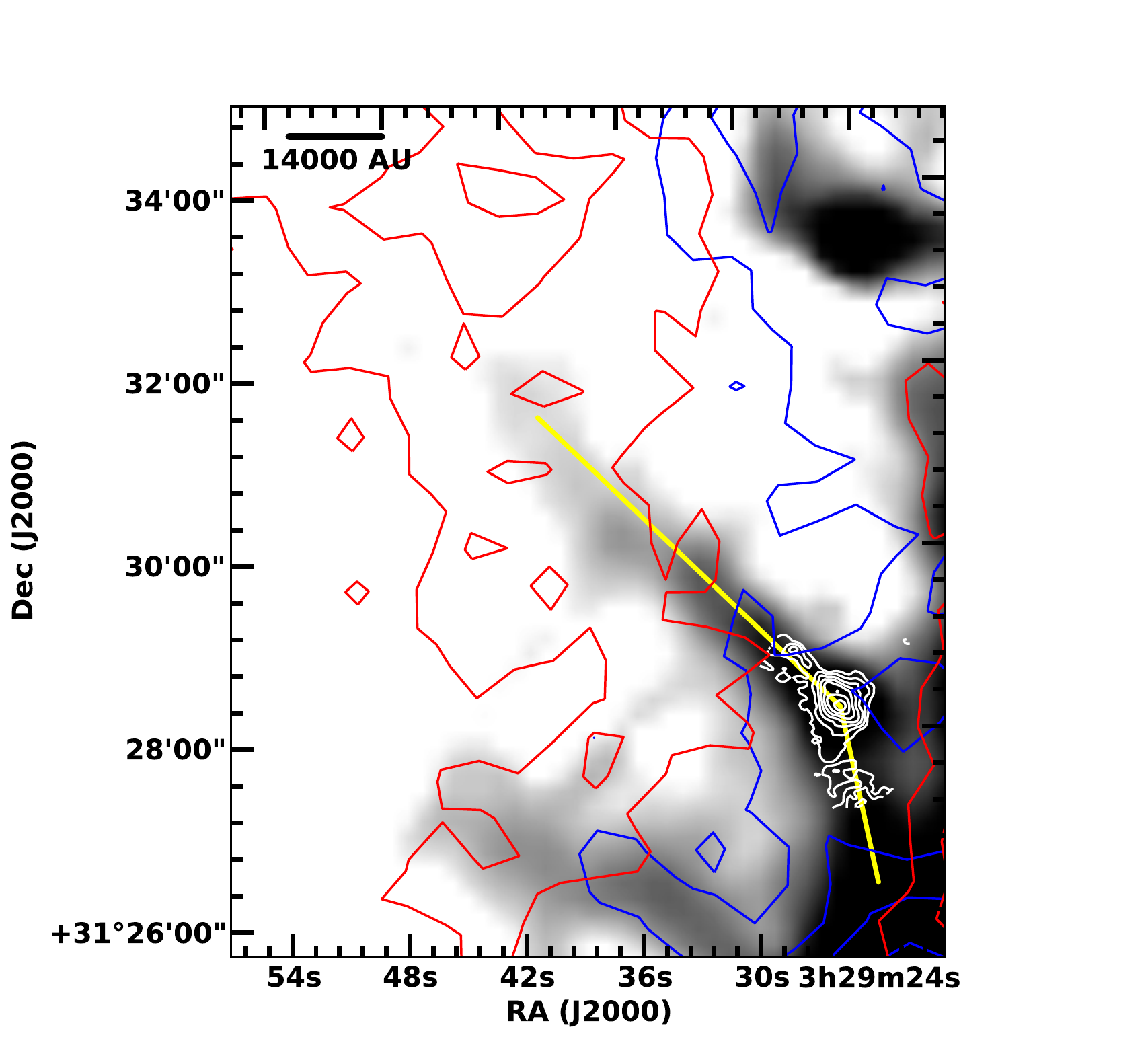}
\caption{\ceo\ and \nthp\ integrated intensity contours over Herschel 250$\mu$m map. The \ceo\ integrated intensity is separated in two velocity bins to better show the distribution of the emission; blue contours show emission in the range 7.2-7.4 \kms, while red contours show emission in the range 7.8-8.0 \kms. White contours show integrated intensity of the CARMA \nthp\ emission from Figure~\ref{fig:perbol58mom0}. Blue and red contours start at 5$\sigma$ and increase in steps of 5$\sigma$. White contours start at 3$\sigma$ and increase in steps of 5$\sigma$. The yellow line shows show the cut for the position-velocity map in Figure~\ref{fig:co_pvmap}.
\label{fig:co_mom0}}
\end{figure}

\begin{figure}
\center
\includegraphics[width=0.5\textwidth]{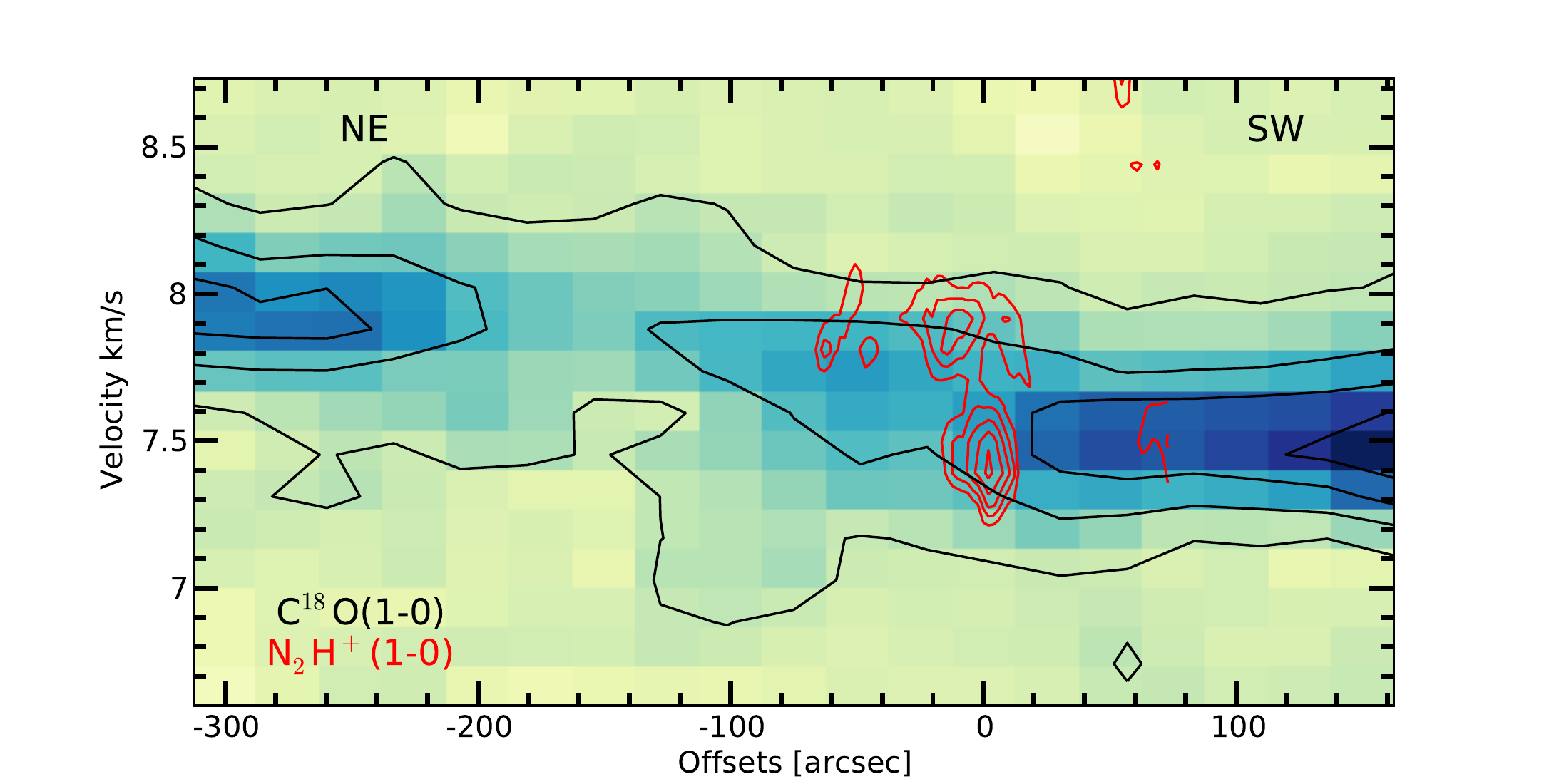}
\caption{\ceo\ position velocity map along the filamentary emission passing through Per-bolo 58 (see yellow line in Figure~\ref{fig:co_mom0}). Red contours correspond to the position velocity map of the \nthp\ emission along the same direction. The spectra at each position in the \ceo\ p-v diagram is averaged over 75" (i.e., the cut is three pixels wide). For the \nthp\ p-v diagram the cut is averaged over the size of the synthesized beam (i.e., 3 pixels). The zero offset corresponds to the continuum peak position of Per-bolo 58. Black contours start at 2.5$\sigma$ and increase in steps of 4$\sigma$. Red contours start at 2.5$\sigma$ and increase in steps of 1.5$\sigma$.
\label{fig:co_pvmap}}
\end{figure}

To further investigate the large scale kinematics of the gas in which Per-bolo 58 is embedded, we examined single-dish \ceo\ observations published by \cite{2003RidgeSurvey} of the northern region of NGC1333, around the location of Per-Bolo 58. Figure~\ref{fig:co_mom0} shows the distribution of the \ceo\ emission around Per-bolo 58 in two velocity ranges 7.2-7.4 \kms\ and 7.8-8.0 \kms, consistent with the velocities of the blue and red velocity components in Figure~\ref{fig:mom0percomp}. We also show the distribution of the \nthp\ emission presented in this work. The yellow line follows the filamentary structure around Per-bolo 58 and indicates the cut that we use for the \ceo\ and \nthp\ position-velocity diagram in Figure~\ref{fig:co_pvmap}. The \ceo\ p-v diagram shows a single velocity peak in the emission for all positions and a velocity gradient along the filament. At the position of Per-bolo 58 the \ceo\ line peaks at velocities more consistent with the blue velocity component defined in this work, while towards the North-East the \ceo\ line peaks at velocities close to the red velocity component. This gradient in velocity is also seen at the filament scale in the  NH$_3$(1,1) observations by \cite{2017FriesenGreen} and in the \nthp\ maps obtained by \cite{2017HacarFibers}. Interestingly, \cite{2017HacarFibers} also identified a filamentary structure with a coherent velocity of 7.4 \kms (a.k.a. fiber, \citealt{2011HacarDense}) passing through the position of Per-bolo 58, and that traces the filamentary emission in the Herschel 250 $\mu$m dust map,  towards the North-East and South of Per-bolo 58 (see Figure~\ref{fig:co_mom0}). At a distance of $\sim$ 10" to the North-East of Per-bolo 58, following the same filamentary structures traced by Herschel, \cite{2017HacarFibers} identified a second fiber at a coherent velocity of 7.8 \kms. These two fiber velocities coincide with the typical velocities of the blue and red velocity components discussed here and found at the envelope scales in this work. It is possible that, due to the close separation in velocity of the two velocity components, the automatic algorithm that \cite{2017HacarFibers} used to fit individual components to the spectra did not detect the two components at the location of Per-bolo 58.\\

Given the velocity structure of the gas surrounding Per-bolo 58 at scales $>0.1$ pc, it seems possible that the two velocity components at small ($\sim$ 1000 au) scales arise from the gas kinematic structure already present at the large scale, i.e from the two filamentary structures, or fibers, at slightly different velocity. We note that it is unlikely that the core formed due to the collision of these two filamentary structures since they both show sub-sonic non-thermal linewidths and the average velocity along these fibers has a dispersion $\leqslant$0.1 \kms \citep{2017HacarFibers}. Instead, a more plausible scenario is that the core slowly accreted from the filaments (or fibers) with slightly different velocities.

\section{Evolutionary state}

One of the goals of this study was to try to determine the evolutionary state of the first core candidate Per-bolo 58, and assess whether this source is a bona fide first core or a very young protostar. Although we find a general agreement regarding the morphology and velocity field of Per-Bolo 58 with the simulation time steps shown here (which are during the main collapse phase, right before and right after the formation of the first core in the simulations), this does not necessarily indicate that Per-bolo 58 is in the first core stage. Similar simulations using a lower resolution also show two velocity components well after protostar formation. Moreover, if the two components are the product of filamentary accretion then, the evolution might be different from the one observed in our simulations. Thus, the appearance of two velocity components is not sufficient for constraining the evolutionary state of this source. \\

Yet, we can set constrains on the evolutionary status of Per-Bolo 58 by comparing it to other sources at known evolutionary stages, where two velocity components have also been detected. \cite{2011TobinComplex} studied a sample of Class 0 protostars using \nthp, with a resolution comparable to that of our observations. They reported one Class 0 protostar, RNO43, in which two velocity components overlap at the center of the core, producing a gradient in velocity somewhat aligned with the outflow direction, as in the case of Per-Bolo 58. There are differences between RNO43 and Per-Bolo 58 that could point towards a younger evolutionary state for the latter. The \nthp\ emission from RNO43 does not peak at the position of the protostar, which seats close to the edge of the \nthp\ emission. In comparison, the \nthp\ total integrated intensity peaks close to the continuum emission in Per-Bolo 58. Furthermore, the distribution of \nthp\ in Per-Bolo 58 is more similar to the previously discussed prestellar core L1544, where the \nthp\ is stronger towards the core center; a sign of a young evolutionary state. When a protostar evolves the temperature of the gas in the inner envelope rises over 20 K at which point the CO that has been locked in the ice mantles of dust grains is released back to the gas phase. The CO reacts with the N$_2$H$^+$ to form HCO$^+$ \citep{2004LeeEvolution}. This produces a drop in the abundance of N$_2$H$^+$ that is observed as a hole, substantial decrease or plateau in the N$_2$H$^+$ emission towards the protostar position, as seen in high-resolution observations of several Class 0 sources (\citealt{2011TobinComplex,2016AnderlProbing}).

Similarly, the linewidth (as fit with a single Gaussian) of the \nthp\ emission in Per-Bolo 58 near the continuum peak is is in between the value observed for the prestellar source L1544 and the protostellar source RNO43. Although there is overlap in the \nthp\ linewidths of prestellar and protostellar sources, there is a trend of smaller linewidths for sources at an earlier evolutionary state \citep{2015HsiehProperties}.\\

In addition, unlike Per-Bolo 58, the \hcop\ p-v map of RNO43 shown in \cite{2011TobinComplex} is consistent with a rotating infalling envelope or Keplerian rotation. The \hcop\ in RNO43 shows clear high-velocity emission tails towards the center of the core, that extend out to about 1 \kms\ from the central velocity. Such velocity structure is not seen in the \hcop\ p-v map of Per-Bolo 58 (Figure~\ref{fig:perbol58pvmap}). Given that these two cores have similar masses, within the uncertainties, the kinematics of the dense gas around these sources suggest that the central object in Per-Bolo 58 has lower mass, and thus younger than the one in RNO43.\\

In \cite{2017MaureiraKinematics} we studied the first core candidate L1451-mm using the same molecular lines and resolution. In contrast to Per-bolo 58, L1451-mm is located in a much more quiescent region of the Perseus molecular cloud complex compared to NGC1333. Also, the L1451-mm core, unlike Per-bolo 58, is at the edge of what seems a filamentary curved structure with a length of $\sim$0.2 pc \citep{2016StormCARMA}. At the envelope scales we studied ($\lesssim$5000 au), L1451-mm shows a single peak profile in all the observed molecular lines; a double peak profile in the \nthp\ is only seen towards the continuum location, consistent with this line being optically thick at this position. A gradient close to perpendicular to the outflow emission was observed in \nhtd. The velocity structure of L1451-mm was interpreted as arising from rotation with infall around a central source of mass $\lesssim$0.06 M$_{\odot}$ \citep{2017MaureiraKinematics}. Although both sources have similar low internal luminosities--- 0.012 L$_{\odot}$ for Per-bolo 58 \citep{2010EnochCandidate}, and $\lesssim$0.016 L$_{\odot}$ for L1451-mm \citep{2011PinedaEnigmatic}--- their kinematics at 1000 au scales are different, which could be indicative of different initial conditions of the gas at the core scales and/or different large scale environments, rather than significantly different evolutionary states. For instance, both sources show similar \nthp\ non-thermal linewidths, which are not observed to be produced by outflow motions and thus likely to be produced by infall, which typically dominates over rotation at these scales. Since infall is expected to increase with the mass of the central object, comparable non-thermal linewidths suggest comparable evolutionary states.

Different initial conditions could also explain why the observed outflows in these sources show contrasting morphologies, despite having similar low characteristic velocities ($\lesssim$ 3\kms). L1451-mm's outflow is not collimated as in the case of Per-Bolo 58; the red and blue lobes in L1451-mm are unresolved with a beam size of $\sim$300 au \citep{2011PinedaEnigmatic}, while Per-bolo 58 red and blue lobes extend up to $\sim$6000 au \citep{2011DunhamDetection}. Although the outflow and envelope kinematics of L1451-mm is more consistent with simulations of first cores formed from the collapse of an isolated core, we can not rule-out that Per-bolo 58 is at a similar evolutionary state. For instance, better outflow collimation can be achieved in simulations if the initial conditions of rotation and magnetic field strength lead to larger magnitudes of the toroidal component of the magnetic field (\citealt{2002TomisakaCollapse,2012SeifriedMagnetic}).\\

A future study that compares L1451-mm and Per-bolo 58 observations with models that include chemical evolution could help to further distinguish the evolutionary state of these sources. For instance, both L1451-mm and Per-bolo 58 show integrated emission of \nhtd, \nthp, \hcn, and \hcop\ peaking at the continuum location. On the other hand the \cs\ integrated emission peaks at the continuum location only for L1451-mm; for Per-bolo 58 the emission decreases towards the continuum. This suggests \cs\ depletion in Per-bolo 58. Given that the abundance of \cs\ is expected to decrease with increasing density during the collapse, and then increase again once the protostar is formed \citep{2004LeeEvolution}, the qualitative difference in the distribution of the \cs\ emission between these two sources could help explain further constrain the relative evolutionary states of these two first core candidates.

\section{Summary}

We present 3mm continuum and molecular line maps of  \nhtd, \nthp, \hcn, \hcop\ and \cs\ with a resolution of 1000 au of the first hydrostatic core candidate Per-Bolo 58. Our results and conclusions can be summarized as follows.

\begin{enumerate}

\item
The line profile of the observed species is composed of two distinct peaks at several locations across the core, separated by 0.4-0.6 \kms. These two peaks are consistent with two different optically thin velocity components rather than the product of self-absorption and high optical depth. The blue velocity component dominates the integrated emission around and at the continuum emission for \nhtd, \nthp\ and \hcn, while the red velocity component dominates the overall core integrated emission for \hcop\ and \cs. The estimated mass of the blue and red velocity components, calculated using the \nthp\ emission under LTE conditions, is about 0.6 and 0.5  M$_{\odot}$, respectively. 

\item
The velocity gradient produced by the transition from the blue to the red velocity component is better aligned with the outflow lobes and the minor axis of the elongated envelope structure, as expected from an approximately edge-on system with an infalling flattened envelope. Both velocity components appear faster at positions near the continuum peak, but they are also high along extended emission tails. The non-thermal linewidths around the continuum emission are subsonic, with \nthp\ showing the narrowest profiles. The red velocity component shows larger non-thermal linewidth, as compared with the blue velocity component.

\item
We compared the morphology and spectra of the \nthp\ with synthetic observations from a MHD simulation that considers the collapse of an isolated magnetized core that is initially perturbed with a turbulent field. We showed synthetic observations from an edge-on view at 0.13 Myr and 0.15 Myr from the start of the collapse, and in both cases, the line profiles were composed of two velocity components, with an emission distribution qualitatively consistent with that of Per-Bolo 58. We conclude that two velocities components can arise in envelopes that have asymmetries in the gas distribution, which in our simulations are produced by turbulence. Thus, we argue that a turbulent and magnetized collapse is a viable explanation for the observed features of this core.

\item
We also discuss the scenario in which the asymmetries in the gas at 1000 au scales are a product of large scale ($\gtrsim$ 0.1 pc) filamentary accretion. The \nthp\ emission shown here is embedded in a large scale \ceo\ filamentary structure along the NE-SW direction, whose velocity towards the NE and SW are consistent with the red velocity and blue components discussed here at the core scales, respectively. Similarly, Per-bolo 58 is close to a region where two large-scale filamentary structures, with velocities that are consistent with the blue and red velocity components, intersect \citep{2017HacarFibers}. Thus, it is also possible that the two velocity components arise due to accretion from filament scales.

\item
As compared with the prestellar source L1544 \citep{2002CaselliMolecular} and the Class 0 source RNO43 \citep{2011TobinComplex} in which two velocity components are also detected, Per-Bolo 58 seems to be at an evolutionary  stage between that of L1544 and RNO43 based on the distribution of the \nthp\ and the kinematics of the \hcop.
The kinematics of Per-bolo 58 are significantly  different from that of the previously studied first core candidate L1451-mm, although both are consistent with a young evolutionary state. These differences, along with their different outflow morphologies could be due to different initial and/or environmental conditions in these cores.

\end{enumerate}

\bibliography{firstcorelibrary}

\end{document}